\documentclass[preprint2]{aastex}

\def\vlsr{$V_{\mbox{\scriptsize LSR}}$}
\def\kms{~km~s$^{-1}$}
\def\kmsyr{~km~s$^{-1}$yr$^{-1}$}
\def\etal{~et\ al.\ }
\def\h2o{H$_{2}$O}

\slugcomment{Submitted to the Astrophysical Journal in 2002.}

\shorttitle{Water masers around RT Vir.}
\shortauthors{H.~Imai \etal}

\begin{document}

\title{The 3-D kinematics of water masers around the semiregular variable RT Virginis}

\author{Hiroshi Imai\altaffilmark{1,2,3}, Katsunori M. Shibata\altaffilmark{2,4}, 
Kevin B. Marvel\altaffilmark{5}, Philip J. Diamond\altaffilmark{6}, 
Tetsuo Sasao\altaffilmark{2}, Makoto Miyoshi\altaffilmark{2}, Makoto Inoue\altaffilmark{4}, 
Victor Migenes\altaffilmark{7}, and Yasuhiro Murata\altaffilmark{8}}

\altaffiltext{1}{Mizusawa Astrogeodynamics Observatory, National Astronomical 
Observatory, Mizusawa, Iwate 023-0861 Japan}
\altaffiltext{2}{VLBI Exploration of Radio Astrometry Project Office, National 
Astronomical Observatory, Mitaka, Tokyo 181-8588, Japan}
\altaffiltext{3}{Joint Institute for VLBI in Europe, Postbus 2, 
7990 AA Dwingeloo, the Netherlands}
\altaffiltext{4}{VLBI Space Observatory Programme Project Office, 
Natinal Astronomical Observatory, Mitaka, Tokyo 181-8588, Japan }
\altaffiltext{5}{American Astronomical Society, 2000 Florida Avenue NW Suite 400, 
Washington DC 20009}
\altaffiltext{6}{MERLIN/VLBI Facilities, Jodrell Bank Observatory, 
University of Manchester, Macclesfield, Cheshire SK 11 9DL, United Kingdom}
\altaffiltext{7}{Department of Astronomy, University of Guanajurato, Apdo Postal 144, 
Guanajuato CP 36000 GTO, Mexico}
\altaffiltext{8}{Institute of Space and Astronautical Science, 3-1-1, Yoshinodai, 
Sagamihara, Kanagawa 229-0022, Japan}

\received{20 November 2002}

\begin{abstract}
We report observations of water masers around the semiregular variable RT Virginis 
(RT Vir), which have been made with the Very Long Baseline Array (VLBA) of the 
National Radio Astronomy Observatory (NRAO) at five epochs, each separated by 
three weeks of time. We detected about 60 maser features at each epoch. Overall, 
61 features, detected at least twice, were tracked by their radial velocities and 
proper motions. The 3-D maser kinematics exhibited a circumstellar envelope 
that is expanding roughly spherically with a velocity of $\simeq$8\kms. 
Asymmetries in both the spatial and velocity distributions of the maser features were 
found in the envelope, but less significant than that found in other semiregular 
variables. Systematic radial-velocity drifts of individual maser features were found 
with amplitudes of $\leq$2\kmsyr. For one maser feature, we found a quadratic 
position shift with time along a straight line on the sky. This apparent motion 
indicates an acceleration with an amplitude of 33\kmsyr, implying the passage of 
a shock wave driven by the stellar pulsation of RT Vir. The acceleration motion is 
likely seen only on the sky plane because of a large velocity gradient formed in 
the accelerating maser region. We estimated the distance to RT Vir to be about 
220 pc on the basis of both the statistical parallax and model-fitting methods for 
the maser kinematics. 
\end{abstract}

\keywords{masers---stars: individual (RT Vir)---stars:evolved, mass-loss, pulsation}

\section{Introduction}

Understanding the dynamics of mass loss flows in circumstellar envelopes of 
evolved stars is one of the most important areas of research in the overall mass 
devolution of stars and the cycling of the interstellar medium. 
Fundamentally, material on the surface of an evolved star gets colder and forms 
dust while it is moving into interstellar space. The newly formed dust is accelerated 
by the stellar radiative pressure and forms an expanding envelope. However, such 
processes are complicated because the related phenomena occur in a dynamically and 
physically unstable region. Maser emission produced from silicon monoxide (SiO), 
water vapor (\h2o), and hydroxyl (OH) molecules are commonly found in many 
circumstellar envelopes (e.g., \citealt{rei81,eli92}). Because of the compactness 
(down to 0.1 AU) of individual maser features, or probably physical gas clumps (e.g., 
\citealt{col92,ima97a}), they are good tracers for investigating the kinematics of 
the envelopes using very long baseline interferometry (VLBI). Multi-epoch 
observations of SiO masers around Mira variables, which are located closer to a stellar 
surface than other masers, using the Very Long Baseline Array (VLBA) suggest that 
circumstellar envelopes of Mira variables are not necessarily spherically symmetric, 
very likely due to anisotropic mass ejection on the stellar surface (e.g., \citealt{dia99}). 
On the other hand, shock waves are also expected simultaneously, which are 
driven by the periodic variation in stellar radiative pressure to the dust in the 
envelopes. These shock waves caused by stellar pulsation are formed near the stellar 
surface and transported to larger distances from the star. 

The gas dynamics of circumstellar envelopes can also be studied with multi-epoch 
observations of water masers with VLBI. The dynamical change of the envelope due 
to stellar pulsation is not so significant ($\sim$10\%) in the near-vicinity of stellar 
surface, a so-called "radio atmosphere" and a "molecular atmosphere" \citep{rei97}. 
In fact, the amplitude of the expansion and contraction motions seen in SiO masers 
in this region is small (10-20\%, \citealt{dia99}). On the other hand, in the outer 
region ($T\leq$1000 K) where water masers are excited, dust formation starts and 
shock waves are enhanced due to dust-induced radiative pressure (e.g., \citealt
{hof95}). A shock wave is expected to produce rapid velocity changes 
(acceleration/deceleration) by up to 10\kms\ in front and in back of it, these will 
be directly detected as radial-velocity drifts and proper motions deviating from a 
constant velocity motion. The present VLBI technology enables us to detect 
such acceleration/deceleration motions with an accuracy of 10 microarcseconds 
($\mu$as) in position and of 0.1\kms\ in radial velocity. Such trials have recently 
started (e.g., \citealt{ish01}, hereafter I01). Water masers are also good tools to 
directly estimate the distances to evolved stars using knowledge of their 3-D velocity 
vectors (radial velocities and proper motions), a method which does not rely on 
the standard distance ladder methods \citep{mar97}. Distance estimation with 
water maser data will be applicable to distance measurements for many evolved 
stars invisible at optical wavelengths after examining the reliability and comparing 
with other methods. 

Here we present the 3-D motions and the radial-velocity drifts of water masers 
associated with the semiregular variable star RT Virginis (RT Vir), which have 
been measured from VLBA data. RT Vir is one of the brightest water maser 
sources (e.g., \citealt{bow94}, hereafter BJ; \citealt{yat94}, hereafter YC; \citealt
{ric99a}a, hereafter RCBY; \citealt{yat00}, hereafter YRGB). The period of stellar 
pulsation of RT Vir has been estimated to be $\simeq$155~d with some 
irregularity (e.g., \citealt {kho85,eto01}). \citet{ima97b} (hereafter Paper 
{\rm I}) found another pulsation period of 375~d using data obtained by the 
American Association of Variable Star Observation (AAVSO). Radial-velocity 
drifts of the water masers, with time, have been found from VLBI observations 
(Paper {\rm I}), however, the origin of the drifts and the true maser kinematics 
are not yet clear because of insufficient angular resolution and a small amount 
of proper motion data (c.f., RCBY; YRGB). 

Section 2 describes the VLBI observations and data reduction. Section 3 summarizes 
the revealed 3-D kinematics of the water masers and the detection of acceleration 
of one maser feature in its proper motion. Section 4 discusses the dynamics of the 
circumstellar envelope of RT Vir, which exhibits acceleration. Measurements of the 
distance to RT Vir are also mentioned there.

\section{Observations and Data Reduction}

The monitoring observations of water masers around RT Vir with the VLBA have 
been made at five epochs during 1998 May--August, with a separation of 3 weeks 
between the successive two epochs. Table \ref{tab:status} summarizes the status 
of the observations. Each of the observations had a duration of 4 hrs including scans 
towards RT Vir and the calibrator 3C~273B for clock offset and complex-bandpass 
calibration. The received signals were recorded with one base-band channel (BBC) 
with a center velocity at \vlsr$=$ 17.0\kms\ and a bandwidth of 4 MHz in dual 
circular polarization mode. The correlated data were processed with 1024 velocity 
channels and a velocity spacing of 0.056\kms\ in each channel at 22.24 GHz. 

The procedures in data reduction using NRAO's AIPS and maser position 
measurement we have applied were almost the same as those presented in 
section 2.2 of \citet{ima00} and section 3.1 of \citet{ima02} (hereafter Paper 
{\rm III}), respectively. In the present work, the complex bandpass characteristics 
were obtained from data on 3C~273B with an uncertainty of less than 1\degr\ in 
phase. For amplitude calibration, we applied a template spectrum of the water 
maser emission obtained from the auto-correlation data. The velocity channel at 
\vlsr$\simeq$17.1\kms\ was selected as reference for fringe fitting and 
self-calibration (see table \ref{tab:status}). The naturally weighted visibility data 
created a synthesized beam of 0.41 mas $\times$ 0.94 mas with a position angle of 
7\degr.6. The detection limit was typically 200 mJy beam$^{-1}$ at a 5-$\sigma$ noise 
level in maps without bright maser emission. The positional accuracy for a maser 
feature was limited typically to 50 $\mu$as mainly because of the extended 
structure of the feature. 

\notetoeditor{Put Table 1 here.}

\section{Results}

\subsection{Proper Motions of Water Maser Features}
\label{sec:pmotion}

\notetoeditor{Put Figures 1.}

\notetoeditor{Put Table 2 here.}

The water masers in RT Vir were spatially well resolved into individual maser features 
with the VLBA synthesized beam. Each maser feature is composed of a compact 
($<$1 AU) bright part and an extended structure. As a result, about 60 maser features 
have been detected at every observation. Table \ref{tab:status} shows the numbers of 
maser features detected at individual epochs. Previous observations with larger 
synthesized beams (Paper {\rm I}; RCBY; YRGB) may not be able to distinguish some of 
individual maser features because of the crowded distribution of maser features in 
the beam. 

We traced the position and radial velocity of each maser feature, which had been 
stable from one epoch to another within 1\kms\ in radial velocity and 0.5 mas in 
position. We also checked the stability of spatial patterns of the maser feature 
(alignment of velocity components or spots, Paper {\rm III}). A total of 60 maser 
features were detected at least twice and their relative proper motions and 
radial-velocity drifts were measured. Table \ref{tab:pmotions} lists measured proper 
motions and radial-velocity drifts of maser features. The estimation of uncertainty 
in proper motion takes into account the apparent sizes of the maser features. Only 36 
of the maser features were detected at least three times, indicating that individual 
maser features turn on and off on a time scale of typically 1--2 months. RCBY and 
YRGB reported maser proper motions of up to 3 mas in 70 days in the year 1996, the 
values were slightly larger than those measured in the present paper (typically 1--2 
mas in 80 days in the year 1998). 

\begin{figure*}
\epsscale{1.5}
\plotone{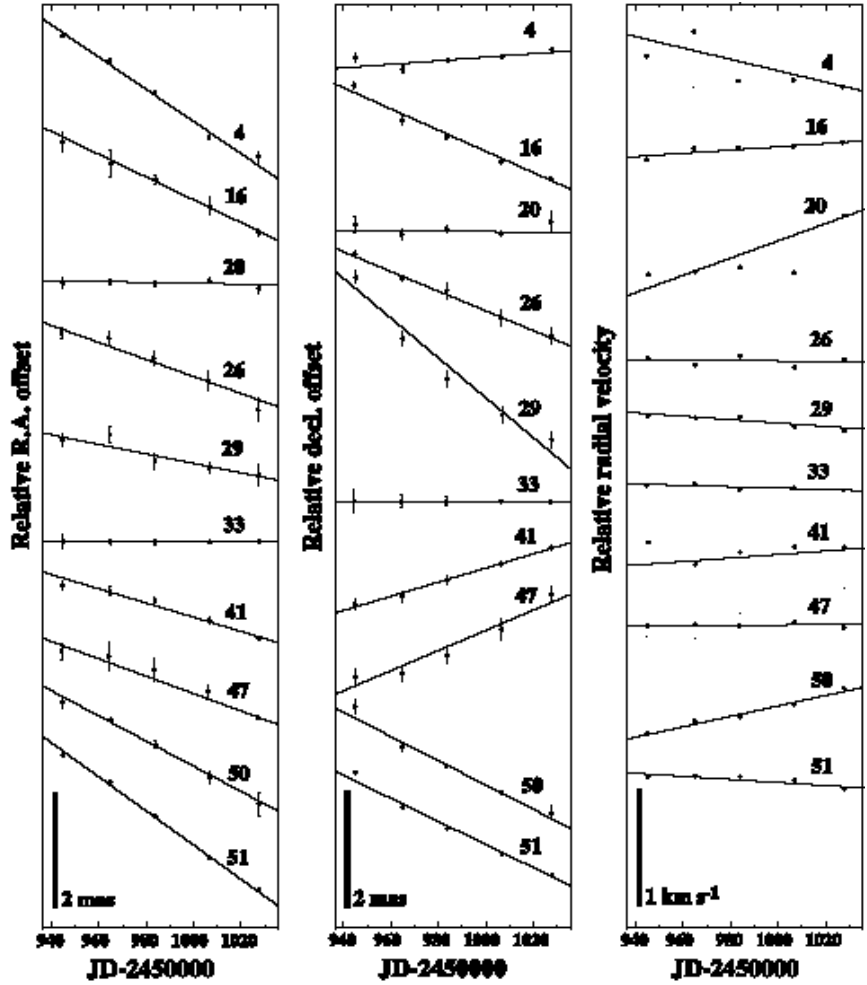}
\caption{A sample of measured proper motions and the Doppler velocity drifts of 
maser features that had been detected in five epochs. The number added after 
"RT Vir:I2002" for each proper motion shows the assigned name. Solid lines 
in plots of proper motions show fit lines assuming constant velocity motions. 
The solid lines in the plots of the Doppler velocity drift show 
fit lines assuming constant acceleration motions. 
\label{fig1}}
\end{figure*}

Figure \ref{fig1} shows changes in positions and radial velocities of maser features 
detected five times. One can find jumps in positions and radial velocities within the 
above stability criteria in some maser features (e.g.\ RT Vir:I2002 {\it 20}). These 
maser features were blended with nearby (closer than 0.5 mas) maser features, for 
which it was difficult to exactly trace their velocities and positions. On the other 
hand, we found not only proper motions that were well fit by a constant-velocity 
motion but also exhibited linear radial-velocity drifts in some features. We also 
found an acceleration motion for a feature in its proper motion, which is described 
in detail in section \ref{sec:acceleration}.

\subsection{The 3-D Kinematics of the RT Vir flow}
\label{sec:3Dkinematics}

\notetoeditor{Put Figures 2, which should appear side-by-side.}

Figure \ref{fig2} shows the angular distribution and the 3-D motions of maser 
features. The features exhibit an elongation in their spatial distribution and 
a radial-velocity gradient in the east--west direction, which have been visible 
in previous observations (\citealt{bow93}; BJ; YC; Paper {\rm I}; RCBY; YRGB). 
The 3-D maser kinematics clearly exhibits an expanding flow, not rotation. The 
kinematic flow looks clearer when including in the diagram the 60 maser 
features detected at least twice than when only including 36 features detected 
at least three times. Because of the crowded distribution of maser features and 
very short lifetimes of maser features, misidentification of feature motions should 
be taken into account especially for the features detected only twice. We made 
further analyses, however, using all of the proper motion data in the present paper. 

\subsubsection{Analysis based on the velocity variance-covariance matrix}
\label{sec:VVCM}

We performed analysis based on diagonalization of the variance--covariance matrix 
(VVCM) of our obtained maser velocity vectors (\citealt{blo00}; I01). The velocity 
dispersions used in the VVCM analysis are the quantities directly determined by 
a proper motion measurement that lacks the absolute position reference. The 
VVCM diagonalization is done by obtaining eigenvectors and eigenvalues for 
the VVCM, which correspond to the kinematic axes of the flow and velocity 
dispersions along the axes, respectively. Thus, the VVCM analysis is an objective 
and model-independent method. The diagonalized VVCM was as follows, 

\[
\left( \begin{array}{ccc}
\sigma_{xx} & \sigma_{yx} & \sigma_{zx} \\
\sigma_{xy} & \sigma_{yy} & \sigma_{zy} \\
\sigma_{xz} & \sigma_{yz} & \sigma_{zz} \\
\end{array} \right)
=
\left( \begin{array}{ccc}
23.84 & 3.09 & 4.19 \\
3.09 & 24.11 & 7.17 \\
4.19 & 7.17 & 21.09 \\
\end{array} \right) 
\]

\begin{equation}
\Rightarrow
\left( \begin{array}{ccc}
15.03 & 0 & 0 \\
0 & 32.79 & 0 \\
0 & 0 & 21.22 \\
\end{array} \right), 
\end{equation}

\noindent
where $\sigma_{ij}=\sigma_{ji}$ is a variance ($i=j$) or a covariance ($i\ne j$) of 
measured maser motions in the $i$- and $j$-axes ($x$, $y$, or $z$) in unit of 
km$^{2}$s$^{-2}$. To estimate uncertainties of the obtained values, we also 
performed a Monte Carlo simulation generating VVCMs with artificial errors 
around the values obtained in the measurement. The eigenvector 
corresponding to the largest eigenvalue (velocity dispersion) had an inclination 
of 24\degr.7$\pm$4\degr.6 with respect to the sky plane and a position angle 
of 35\degr.8$\pm$14\degr.9, which is roughly parallel to both of the directions 
of the elongation and the velocity gradient mentioned above. 
This implies the bipolarity of the RT Vir flow, but is not so significant, since the 
velocity dispersion is roughly equal in all directions; a ratio of the eigenvalue was 
2.47:~1.67:~1 (c.f., the ratio of 6.1:~2.0:~1 in R Crt, I01). 

\subsubsection{Model fitting for the maser kinematics}
\label{sec:model-fitting}

We also made a least-squares model-fitting analysis assuming a spherically 
expanding flow model, details of which have already been described in \citet{ima00} 
(c.f. \citealt{gwi92}). We used weights proportional to the square of the accuracy 
of a measured proper motion.

\begin{figure*}
\vspace*{-1cm}
\epsscale{1.7}
\plotone{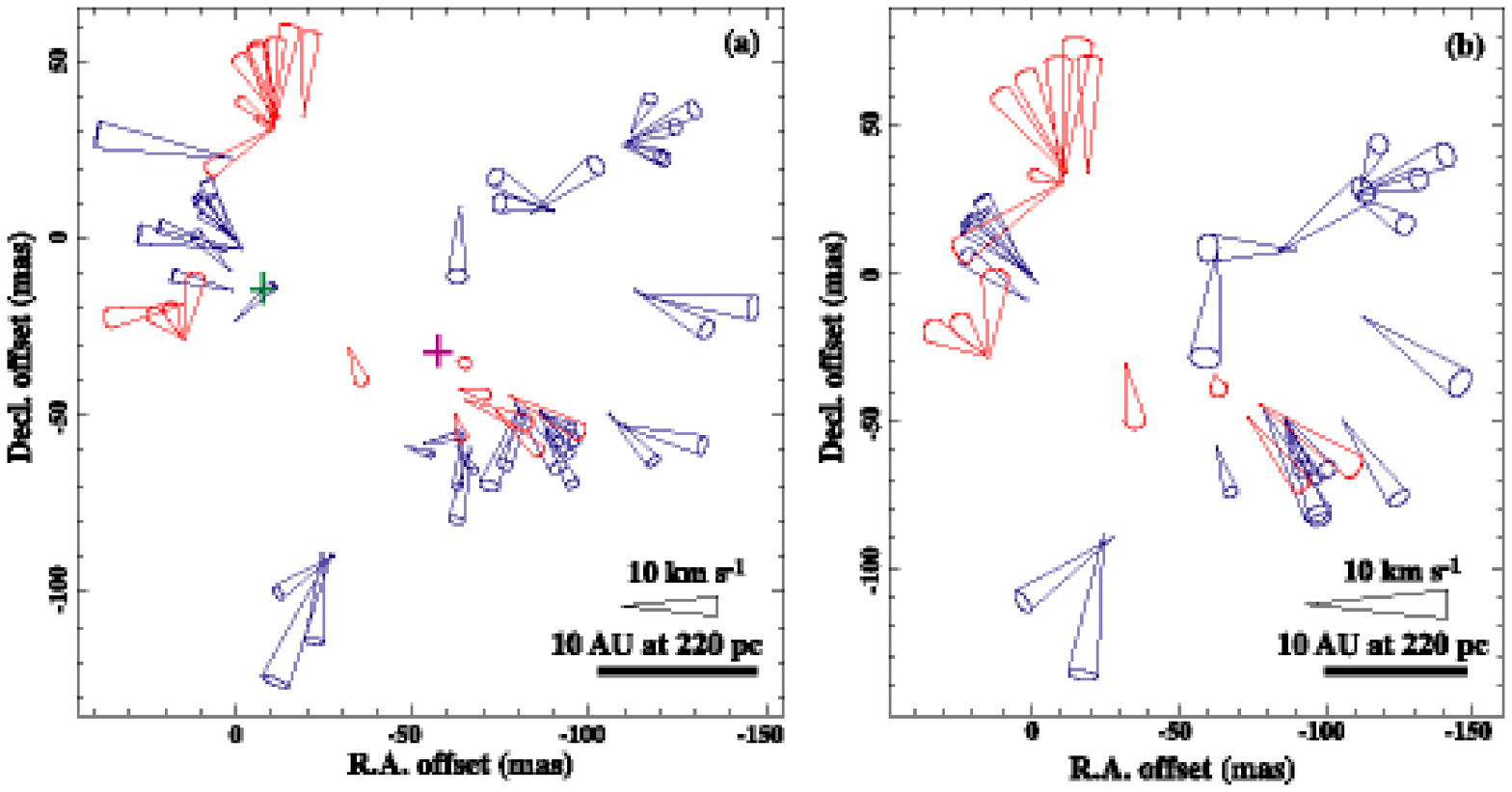}
\caption{The 3-D velocity field of water maser features around RT Vir. 
A 3-D velocity vector of a feature is indicated by a cone. The mean velocity vector 
of the maser features was subtracted from each of the observed velocity vectors. 
(a): the 3-D motions of 61 maser features, each of which had been detected at least 
twice, are shown. The plus symbols indicate the locations of the star, estimated by 
the model fitting using proper motions exhibiting both expansion and infall (pink) 
and only expansion (green). (b): Same as (a) but for 36 maser features, each of which 
had been detected at three epochs.
\label{fig2}}

\clearpage 

\epsscale{1.0}
\plotone{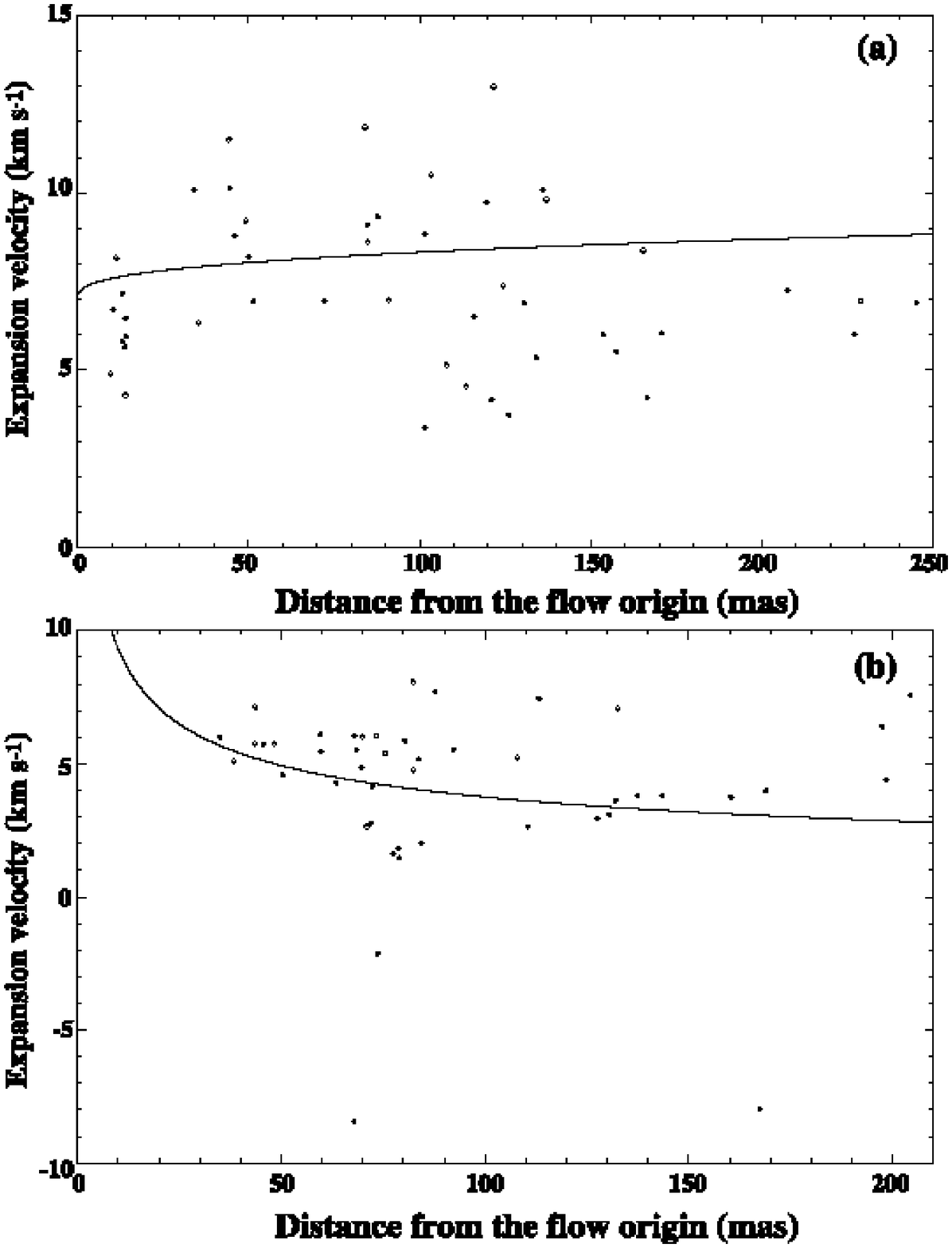}
\caption{The expansion velocity of a maser feature against the distance from 
the position of the central star. These parameters were estimated after the 
model fitting. The expansion velocity was calculated by using the equation 
shown by \citet{ima00}. A solid line shows the velocity field as a power-law 
function with respect to the distance from the star. (a): For the case when 
excluding infalling features, having negative expansion velocities. (b): 
For the case when including all maser features within a reasonable distance 
from the star ($<$ 300 mas, corresponding to 66 AU at 220 pc). 
\label{fig3}}
\end{figure*}

First, we adopted radial expansion motions of masers with independent speeds 
and estimated only a systemic bulk motion, or the motion of the star, and a 
location of the star as free parameters. Then, we also estimated a distance to 
RT Vir and the velocity field of the radial expansion. In the present paper, we 
assumed the speed of the radial expansion of a maser feature {\it i}, 
$V_{\mbox{\scriptsize exp}}(i)$, as a function of the distance to a maser feature 
from the origin of the outflow, $r_{i}$, which is expressed as 

\begin{equation}
\label{eq:exp-speed}
V_{\mbox{\scriptsize exp}}(i)\;=\;V_{0}+V_{1}\left(\frac{r_{i}}{r_{0}}\right)^{\alpha},
\end{equation}

\noindent
where $V_{0}$ is the intrinsic velocity at the stellar surface, $V_{1}$ the velocity at 
a unit distance $r_{0}$, $\alpha$ the power-law index indicating the apparent 
acceleration of the flow. We applied the fitting technique step-by-step, excluding 
maser features with unreliably large distances from the outflow origin ($>$0\arcsec.3 
or 66 AU at 220 pc). Table \ref{tab:RTVir-model-fit} shows the best solutions obtained 
in the analyses. Figure \ref{fig3} shows the distribution of features with expanding 
velocities obtained using equation (10) of \citet{ima00} versus distances from the 
estimated position of the central star.

\notetoeditor{Put Figures 3.}

Through this procedure, we found that several maser features had negative 
expansion velocities or infall motions towards the star (see also figure \ref{fig2}). 
Although some of the features were detected at least at three epochs, it is still 
unclear whether they were tracing actual physical motions. When including 
these infalling maser features, the best-fit function indicates a slow 
($V_{\mbox{\scriptsize exp}}<$5\kms) expansion of the flow with marginal 
deceleration (figure \ref{fig3}b). These are inconsistent with the 
suggestion that the RT Vir flow has an expansion velocity 
$V_{\mbox{\scriptsize exp}}\simeq$10\kms\ in the water maser region and 
radial acceleration toward the outer OH maser region (RCBY). The estimated 
position of the star is roughly at the center of the whole maser distribution 
but slightly biased toward clusters of the blue-shifted maser features located 
at the south-west of the whole region. The estimated position of the expansion 
center or the star is located close to the center of the "ring" found in the maser 
distribution. No maser feature is located within 30 mas ($~$7 AU) from the star, 
indicating a "quenching zone" of water maser emission (YC).

On the other hand, when excluding the maser features exhibiting infall, a 
higher expansion velocity $V_{\mbox{\scriptsize exp}}\simeq$8\kms\ was 
obtained. The estimated location of the star is very close to blue-shifted clusters 
of maser features. We could not find converging solutions in which the stellar 
position is close to the center of the ring mentioned above. Thus, the spatial 
distribution of the masers is expected to be significantly asymmetric (see also 
figure \ref{fig2}). This is consistent with the estimation by BJ. Maser features 
apparently spread out to 200 mas ($\sim$45 AU) from the star (see also figure 
\ref{fig3}), which is larger by a factor of 2--3 than that previously estimated 
(\citealt{bow93}; BJ; YC). The size of the maser-quenching zone is reduced to 
10 mas ($\sim$2 AU). The estimated systemic radial velocity of the star also 
has an offset of $\sim$3\kms\ from the value adopted by above previous papers. 

\notetoeditor{Put Table 3.}

\notetoeditor{Put Figures 4 and 5.}

\begin{figure*}
\epsscale{1.7}
\plotone{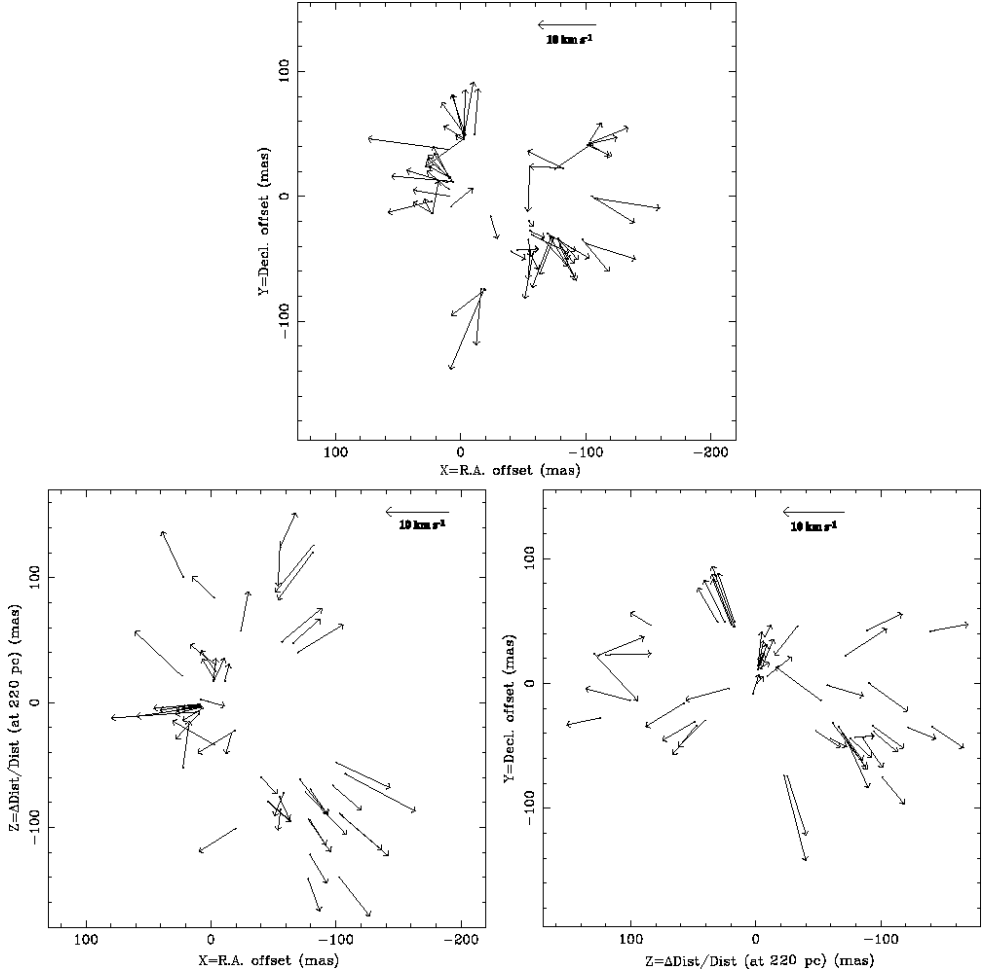}
\caption{Estimated 3-D positions and motions of water maser 
features in RT Vir. The positions and motions are with respect to those of 
the outflow's origin determined by a model fit. The position of the arrow 
indicates that of a maser feature. The direction and length of the arrow 
indicates the direction and the magnitude of the maser motion, respectively. 
{\it Top}: Front view (\mbox{$XY$}-plane) for the positions and motions plot. 
{\it Bottom left}: Top view (\mbox{$XZ$}-plane). {\it Bottom right}: East-side 
view (\mbox{$ZY$}-plane).
\label{fig4}}
\end{figure*}

Figure \ref{fig4} shows the estimated 3-D spatio-kinematics of the water masers 
projected in three directions. Most of the maser features exhibit radial expansion 
without rotation around the star at the diagram origin. Contrary to the asymmetric 
spatial distribution of maser features, asymmetry in the velocity field is not seen; 
maser features seem to have only a bias in the spatial distribution. Thus, the 
kinematics of the RT Vir flow is well expressed by a radially expanding flow (c.f., 
Paper {\rm I}; RCBY; YRGB) and the radial-velocity gradient in the east-west 
direction should be due to the bipolarity of the flow but with weak collimation. 

\subsection{Acceleration motions found in water maser features}
\label{sec:acceleration}

\notetoeditor{Put Figures 6 and 7}

We have measured radial-velocity drifts of individual maser features by measuring 
a radial velocity at the brightness peak of each maser feature in the same manner 
as that in Paper {\rm III}. Because we measured the velocities with accuracy better 
than a velocity channel spacing of 0.056\kms, the uncertainties of the velocity drift 
rates were calculated by assuming a measurement error to be equal to this value. 
The uncertainties were as small as 0.3\kmsyr\ in the best case. Table \ref{tab:pmotions} 
gives the measured radial velocity drifts of maser features. Figure \ref{fig1} presents 
time variations in radial velocities of individual maser features with longer lifetimes. 

The changes in radial velocity are usually less than the velocity widths of maser 
features (0.5--2.0\kms, see table \ref{tab:pmotions}) and likely to be affected by 
temporal changes in spatial and velocity structures of maser features with time. 
\citet{gwi94} pointed out that the observed line width is likely affected by the 
subsonic bulk motions within the feature ($\leq$0.5\kms) and the different 
hyperfine transitions of the \h2o\ line (but negligible). Because the two effects 
will also affect the radial-velocity drift measurements, the velocity variations 
larger than 0.5\kms\ can be taken into account as possible accelerations. 

\begin{figure*}
\epsscale{1.0}
\plotone{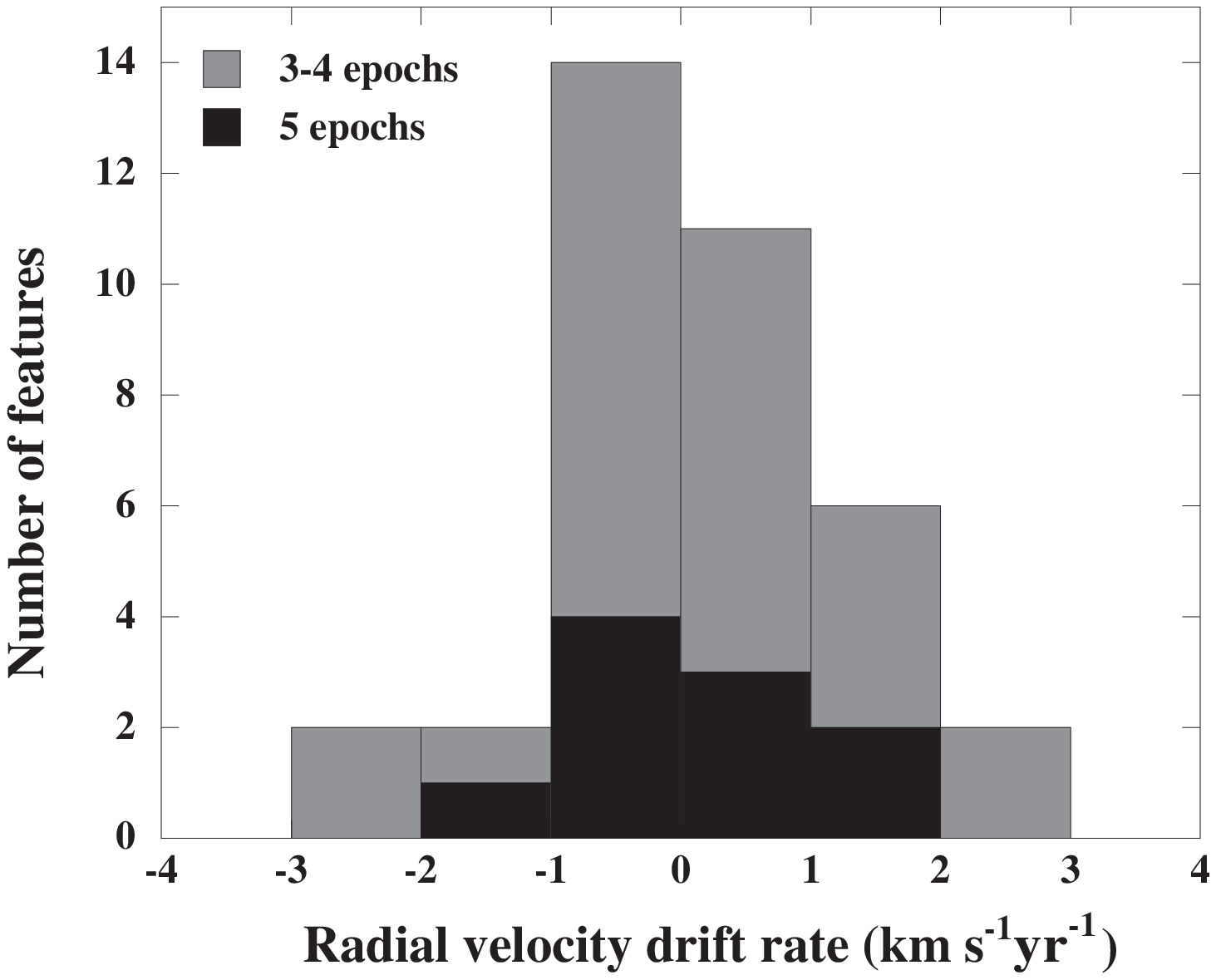}
\caption{Histogram of drift rates of maser features' radial velocities. Maser features 
are divided into two groups, features detected at all epochs and those detected at 
3 or 4 epochs. Maser features detected only at two epochs have been excluded.
\label{fig5}}
%
%
\epsscale{1.8}
\plotone{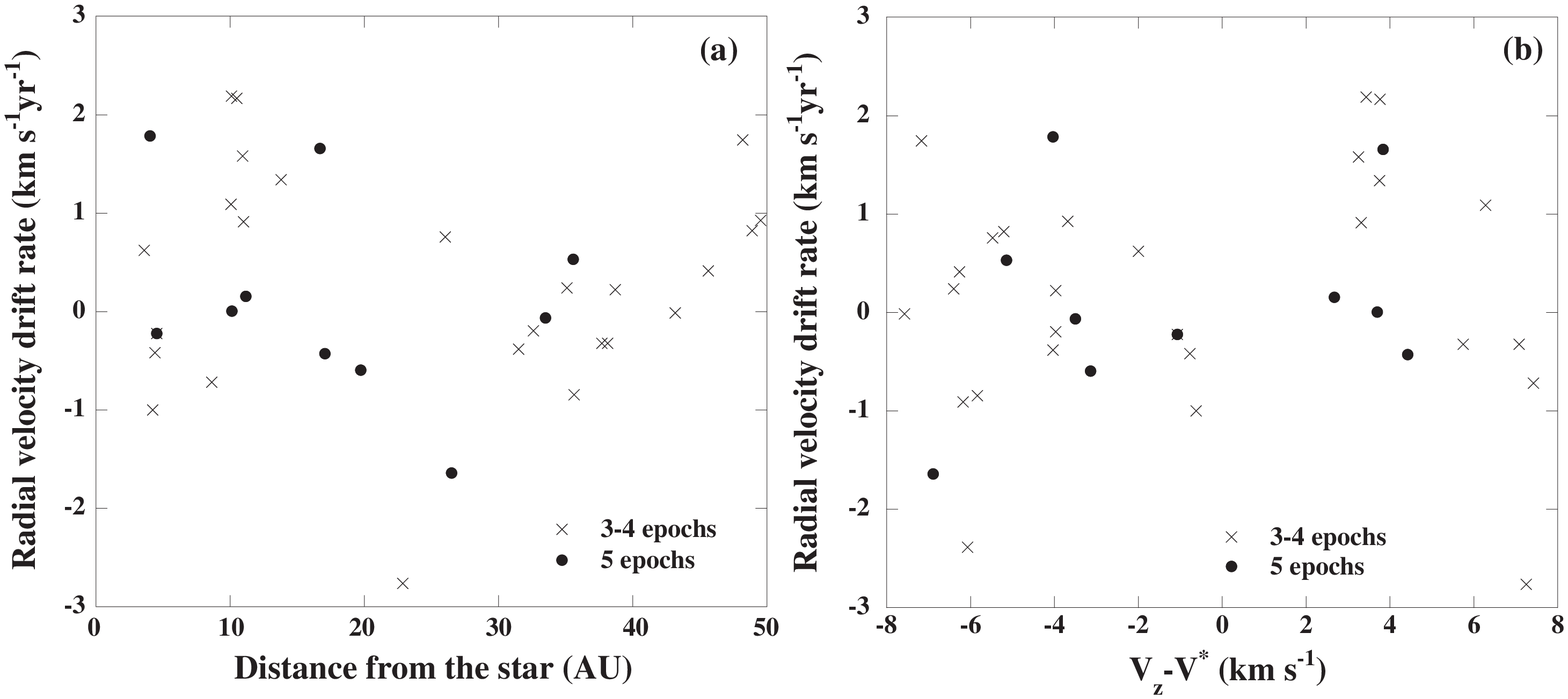}
\caption{Dependence of the radial-velocity drift of a water maser feature on its 
location and radial velocity. (a): Against distance from the star (b) Against radial 
velocity with respect to the assumed stellar velocity (\vlsr$=$18.0\kms). 
No correlation is found between radial-velocity drifts and locations/radial velocities.
\label{fig6}}
\end{figure*}

Nevertheless, some of the maser features changed their velocities with time at 
constant rates, or constant accelerations. Figure \ref{fig5} shows the histogram 
of the measured radial-velocity rates. The drift rates of these features are equal 
to or smaller than 1\kmsyr\ and quite similar to those that have been commonly 
detected for 15 yrs with single-dish observations \citep{lek99}. Maser features 
with larger drift rates ($>$2\kmsyr) exhibit jumps in velocities, which are likely 
due to blending of a few very close by maser features as mentioned in section 
\ref{sec:pmotion}. Such large velocity drifts were also found in some maser 
features detected in less than four epochs. Thus, radial-velocity drifts 
around 1\kmsyr\ come from single maser features and may indicate 
acceleration/deceleration of the features. Previous VLBI observations (Paper 
{\rm I}; I01) have also detected such systematic radial-velocity drifts in a small 
number of maser features. We did not find, however, a correlation between 
radial-velocity and acceleration (see figure \ref{fig6}). 

\begin{figure}
\epsscale{1.0}
\plotone{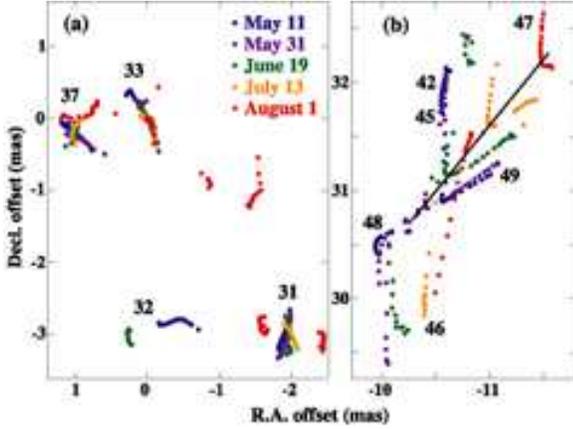}
\caption{Spatial structures of maser features and their time variation in the two 
selected fields. The number added after "RT Vir:I2002" for each maser feature 
shows the assigned name. Each of the filled circles shows a velocity component 
(maser spot). Radial velocities of spots change by 0.056\kms\ from one spot 
to the adjacent spot. (a): Maser features around the map origin. The 
position-reference feature, RT Vir: I2002 {\it 33}, is spatially fixed at the map 
origin to make measurements of maser proper motions, but will move from 
south to north so that it expands with respect to the star. (b): Maser features 
exhibiting acceleration motions. The maser feature on a black line, RT Vir:I2002 
{\it 47}, shows a clear constant acceleration motion and the development of its 
V-shaped structure. The maser feature had kept its brightness peak close to the 
crease of the V-shaped structure. 
\label{fig7}}
\end{figure}

We have also looked for maser features exhibiting acceleration/deceleration 
motions in their proper motions. Similar to measurements of the radial-velocity 
drifts, we traced positions of brightest peaks of individual maser features. 
The method to define the feature position was described in Paper {\rm III}. 
Note that flux-weighted positions (e.g., \citealt{mar97}) are often affected by 
extended and weak velocity components. Figure \ref{fig7}a shows the time 
variation in spatial structures of the position-reference feature (RT Vir 2002I: 
{\it 33}) and the adjacent features. We find that the reference feature has 
changed in spatial structure from one epoch to another. Most of the maser 
features had similar variation in their internal structures, which makes it more 
difficult to judge whether an apparent acceleration motion really reflects a true 
one or a physical motion of the maser cloud. Because VLBI observations lose the 
absolute position coordinate, we can not calibrate the change in spatial 
structure of the reference feature for the proper motion measurements. 
Assuming that this change is coming from the subsonic bulk motions within 
the feature mentioned above, an uncertainty of $\sim$0.5\kms\ (0.5 
mas~yr$^{-1}$) likely contaminates the measured relative proper motions 
of other maser features. 

\begin{figure}
\plotone{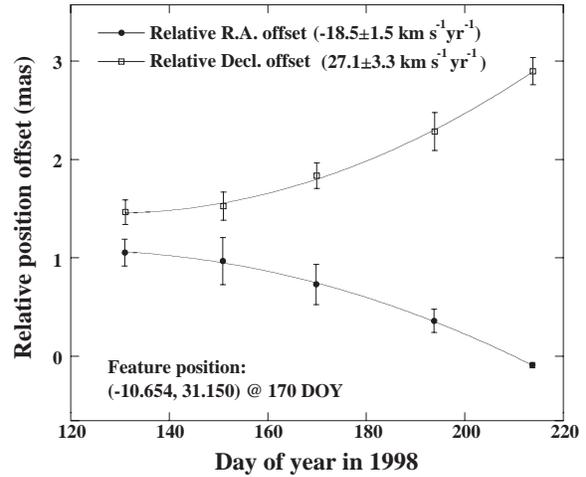}
\caption{The temporal position variation of the maser feature RT Vir:I2002 {\it 47}, 
which apparently shows an acceleration motion. Solid lines indicate fits to the 
position variation in R.A.\ and decl.\ directions assuming a constant acceleration 
motion. The positions of the maser feature at individual epochs were defined 
to be at the brightness peak among the maser spots in the maser feature and 
measured with uncertainties less than 100 $\mu$as. A vertical bar with the 
feature is an extension size of the feature, or a distribution size of the spots in 
the feature. 
\label{fig8}}
\end{figure}

Nevertheless, we found a maser feature candidate that may exhibit a true 
acceleration motion, in which the feature moved by a distance much larger than 
the size of the feature itself. Figure \ref{fig7}b shows the time variation of the 
candidate, RT Vir 2002I: {\it 47}, and the adjacent features. Figure \ref{fig8} 
shows the trajectory of the candidate and a model, which adopts a constant 
acceleration and is fitted well to the observed trajectory. The acceleration vector 
([$-$18.5, 27.1] in units of \kmsyr) has a position angle of ($-$34\degr, which is 
roughly parallel to the mean proper motion vector ([$-$2.2, 10.4] in unit of \kms, 
subtracting the mean proper motion shown in table \ref{tab:RTVir-model-fit}) 
with a position angle of $-$12\degr\ and the position vector ([$-$2, 46] in unit of 
mas) with a position angle of $-$2\degr. The measured acceleration motion 
vector is a relative one with respect to that of the reference feature, or we are 
measuring it in the frame in which the reference feature has a constant velocity 
motion. Expecting spherically symmetric acceleration motions in the features 
RT Vir 2002I: {\it 47} and {\it 33} with respect to the central star, the relative 
acceleration vector would be closer to the proper motion vector of the feature 
RT Vir 2002I: {\it 47}. Adopting this acceleration, the feature increased in speed 
from 1\kms\ to 16\kms. 

Note that the apparent acceleration of the feature was seen only on the sky 
plane ($\simeq$33\kmsyr), but not along the line-of-sight ($\leq$0.1\kmsyr ). 
Other maser features also exhibited apparently rapid acceleration motions 
on the sky plane but much smaller ones along the line-of-sight as mentioned 
above ($\leq$3\kmsyr). 

\section{Discussion}

\subsection{Stellar position in the \h2o\ maser distribution}

The stellar position of RT Vir in the circmstellar envelope has been estimated 
in several manners based on the maser maps or position-velocity diagrams. 
All of the results suggest that the star is close to the center of the \h2o\ 
maser distribution or the middle between the blue-shifted and red-shifted 
clusters of maser features (e.g., \citealt{bow93}; YRGB). YC and \citet{bai02} 
found a "quenching zone", within which no maser emission is found. \citet
{rei90} found that a radio continuum source from the semiregular variable 
W~Hya is located at the center of a "ring" found in the \h2o\ maser distribution. 

On the other hand, the dynamical center of the expanding flow estimated 
in the present paper has a large offset ($\sim$40 mas) over the uncertainty 
in the model fitting (see table \ref{tab:RTVir-model-fit}). The stellar position 
can be examined by simultaneously making proper motion measurements 
such as those in the present paper and observations of the radio continuum 
source of RT Vir such as that made by \citet{rei90}. It is expected that the 
asymmetry of the 3-D distribution of maser features (see figure \ref{fig4}) 
derived a bias in the model fitting. Long-term observations, over several years, 
will reduce such distribution asymmetry.

\subsection{Distance to RT Vir}

The estimation of the distance to RT Vir has been made with several methods: 
a trigonometric parallax with the {\it HIPPARCOS} satellite (140 pc) and bolometric 
distance assuming a P-L relation (120--360 pc, e.q., BJ; YC; \citealt{yua99}). 
In the present paper, we made the distance estimation on the basis of the 3-D 
kinematics of the RT Vir water masers using both, the statistical parallax and the 
model-fitting methods. As described in section \ref{sec:VVCM}, the feasibility of 
these methods was examined by the objective analysis using VVCM, which 
shows no rotation or other peculiar motion in the velocity field (see also figure 
\ref{fig5}). In the model-fitting method, the detail of which has already been 
described in section \ref{sec:model-fitting} and table \ref{tab:RTVir-model-fit}, 
we obtained distance values of 226$\pm$16 pc in the analysis using only maser 
features exhibiting expansion. 

When including infalling features in the analysis, the distance value was 
85$\pm$12 pc. Then we obtained an expansion velocity 
$V_{\mbox{\scriptsize exp}}\sim$5\kms (see figure \ref{fig3}), which is smaller 
than the value obtained by RCBY (V$_{exp}\simeq$10\kms). The results of RCBY 
suggested that the circumstellar envelope is accelerated gradually toward the 
outside, from 2-3\kms\ in the SiO maser region closest to the star to 
20--30\kms\ in the OH maser region most distant from the star. It is expected 
that RCBY's results reflected a "steady" flow of RT Vir. Because, as expected 
in theoretical models (e.g.\ \citealt{hof95}), infall motions occur as a result of 
a time-variable phenomenon, or shock wave, the infalling features are needed 
to be excluded in the model assuming the steady flow. Thus, the above 
distance value is concluded to be underestimated. 

On the other hand, the statistical-parallax method \citep{gen81} gave distance 
values of 237$\pm$29 pc (232$\pm$27 pc and 243$\pm$29 pc for the 
$\mu_{x}$--$V_{z}$ and the $\mu_{y}$--$V_{z}$ data, respectively) and 
214$\pm$28 pc (211$\pm$27 pc and 216$\pm$28 pc for the 
$\mu_{x}$--$V_{z}$ and $\mu_{y}$--$V_{z}$ data, respectively) when using all 
3-D motion data and only the data used in the former model fitting, respectively. 
The distance values obtained in the former model-fitting and the statistical 
parallax methods are quite consistent with each other, and the weighted-mean 
distance value of 220$\pm$30 pc is adopted for RT Vir in this paper. 

Note that the statistical parallax methods used only maser features with measured 
proper motions. Adopting a radial velocity dispersion of 66 maser features detected 
in the first epoch when the largest number of features were detected, a distance 
value of 279$\pm$32 pc was obtained. Thus, the radial velocity dispersion tends 
to have the larger value, with respect to the proper motion dispersion, when 
including maser features without measured proper motions. This implies that 
maser features without measured proper motions have the different kinematics 
from that of maser features with the proper motions. The distance estimation in 
this paper was the first successful one for a water maser source around a low-mass 
Mira-type star next to those around supergiants exhibiting many long-lived water 
maser features \citep{mar97}. 

\subsection{Asymmetry of the mass-loss process in semiregular variables}

The morphology and the kinematics of circumstellar envelopes (or mass-losing 
flows) of evolved stars, which are traced by the water maser kinematics, 
sometimes exhibit clear bipolarity or significant asymmetry. Even if taking into 
account the limited physical conditions for maser excitation, such asymmetry 
is evident throughout the revealed 3-D maser kinematics. At the end of the 
AGB phase in stellar evolution, such asymmetry will be tightly related to formation 
of elongated planetary nebulae, some of which are created by stellar jets (e.g., 
\citealt{ima02b}). The asymmetry, however, has been also found in semiregular 
variables (e.g., \citealt{bow93}; BJ; YC; I01), which has been considered to be at 
the early-AGB phase. The present and previous works (e.g., \citealt{mar97}: I01; 
\citealt{yat00}) have proven that radial-velocity gradients seen in water maser 
distributions are due to the bipolarity, not the rotation of mass-losing flows. 
On the other hand, many Mira and semiregular variables, including RT Vir, do 
not clearly show the flow bipolarity yet, having major/minor axis ratios of 
$\leq$2. It is still difficult to discuss the relation among the asymmetry, 
mass-loss rate, radius, and expansion velocity of a mass-losing flow due mainly 
to the limited number of sample stars available for data of the 3-D maser 
kinematics and for accurate distances. In this paper, we speculate about 
the weak asymmetry of the RT Vir flow. 

We re-estimated a mass-loss rate for RT Vir adopting the described 
spherically-expanding flow model. The density of material (hydrogen molecules) 
in the flow is determined from predictions by a maser excitation theory as  
10$^{6}$ cm$^{-3}\; \leq\; n_{\mbox{\scriptsize H$_{2}$}}$ 
$\leq$ 10$^{9}$ cm$^{-3}$ (e.g., \citealt{coo85,eli92}). Adopting the radius of the 
flow in the water maser region, $r\simeq$44 AU, at $d\simeq$220 pc 
(distribution size of $\simeq$200 mas), an expansion velocity, 
$V_{\mbox{\scriptsize exp}}\simeq$8\kms, we obtained a mass-loss rate of 
$\dot{M}\sim$ 1.5$\times$10$^{-7}$
($n_{\mbox{\scriptsize H$_{2}$}}$/10$^{6}$ cm$^{-3}$) $M_{\sun}$ yr$^{-1}$, 
whose lower limit is roughly consistent with that estimated by BJ. Adopting 
the relation between a radius of the water maser distribution and a mass-loss 
rate found by \citet{coo85} and \citet{lan87}, the mass-loss rate of RT Vir 
should be much larger than this value (c.f., 
3$\times$10$^{-6}$ $M_{\sun}$ yr$^{-1}$, BJ). 

On the other hand, Mira-type stars with relatively large mass-loss rates and 
larger flow radii have symmetric kinematics (e.g., BJ). In OH/IR stars with 
larger mass-loss rates, 1612-MHz OH masers exhibit larger and symmetric 
distributions (e.g., \citealt{cha86}). It is considered that the intrinsic bipolarity 
or asymmetry of a mass-losing flow is obscured by the growing envelope 
as the mass-loss rate increases. In this case, RT Vir will be in the transition 
during an increase in its mass-loss rate. We have to take into account, however, 
cases of supergiants, which have much higher mass-loss rates but most of 
which exhibit significant asymmetry. 

Note that the expansion velocity and the radius of a mass-losing flow determined 
seem to be time dependent. Water masers around R Aql are one of the examples 
(BJ). The bipolarity of the maser distribution was significant after the maximum of 
the light curve and vice versa. A similar variation in the maser distribution has 
been confirmed for RT Vir, but when a stellar pulsation period of 375~d is 
adopted (Paper {\rm I}). Adopting this tendency, water masers around RT Vir 
were observed around the light maximum, in which the maser distribution was 
larger than those observed with the VLA (\citealt{bow93}; BJ) but smaller than 
those observed with the Japanese VLBI Network (J-Net, Paper {\rm I}). If we 
could measure maser proper motions with the J-Net data, the bipolarity would 
be exhibited more clearly. The time dependence of the bipolarity should be 
examined with monitoring observations for longer than a few cycles of the 
pulsation period. 

\subsection{Origin of apparent acceleration motions}

Long-term monitoring observations have showed that there are apparent 
radial-velocity drifts of water maser features that have occurred 
simultaneously and exhibited the same drift direction 
(acceleration/deceleration) among several maser features \citep{lek99}. 
This implies a change of the velocity field in the mass-losing flow due to 
the stellar pulsation as suggested above. In the case of RT Vir, such 
acceleration/deceleration observed is likely excited by stellar pulsation 
with a period of $\simeq$375~d \citep{ima97b} rather than $\simeq$155~d 
(e.g., \citealt {kho85,eto01}). Such systematic velocity drifts may be 
detectable only with monitoring observations spanning at least one year. 
It is because there are small changes in the drift rates on this time scale 
and because it is sometimes difficult to remove data exhibiting 
velocity jumps due to blended features having intensity variation with time. 
In fact, in short duration ($<$ a few months) the velocity drifts have not 
exhibited the same drift direction among maser features in both the previous 
and the present VLBI observations (see figure \ref{fig6}, also Paper {\rm I}; I01). 

On the other hand, theoretical models adopting shock waves driven by stellar 
pulsation predict that a shock wave creates acceleration/deceleration of a 
mass-losing flow and generates rapid velocity changes during the passage of 
the shock (a few months) (e.g., \citealt{hof95}). The velocity changes are 
expected to be $\sim$10\kms\ and $\sim$2\kms\ at the inner ($r\sim$5 AU) 
and the outer ($r\sim$50 AU) region of water maser excitation, respectively. 
The acceleration motion observed in a feature's proper motion, as mentioned in 
section \ref{sec:acceleration}, is quite consistent with the theoretical prediction. 

It has been a long-term issue whether change in the location of a maser feature 
is tracing actual physical movement of the gas in a clump rather than some kind 
of non-kinematic effect, such as traveling excitation phenomena or chance 
realignment of coherency paths through the masing gas. Recent observations, 
however, have strongly suggested the former in hydroxyl and water masers; 
all of the observations have shown that individual maser features persist in both 
their spatial and radial-velocity patterns throughout their motion for distances 
much larger than their own sizes (e.g., \citealt{blo96, tor01a}a, b; Paper {\rm III}). 
The maser feature (RT Vir I2002:{\it 47}, see figure \ref{fig7} and \ref{fig8}) 
exhibited a large acceleration on the sky, but simultaneously a large position 
drift of 2 mas, a stable radial velocity (variation less than 0.2\kms). This feature 
also exhibited a persistent V-shaped pattern formed by a cluster of maser spots, 
which have radial velocities successively changing by 0.056\kms\ from one spot 
to another. These strongly support that this maser feature traces the movement 
of a gas clump throughout five observation epochs. 

The larger acceleration on the sky plane (over 10\kms) than that in the 
line-of-sight ($\simeq$1\kmsyr) is likely due to the beaming effect of maser 
radiation. Taking into account the hydrodynamical treatment of a maser 
clump, an apparent acceleration motion of a brightness peak in the maser clump 
is expressed as, $\slantfrac{\mbox{d}V_{r}}{\mbox{d}t}$ $\approx$ 
$V_{r}\cdot \slantfrac{\partial V_{r}}{\partial r}$, where $V_{r}$ and $r$ are the 
velocity field as a function of distance from the star, $r$, and the location 
in the clump, respectively. Assuming a velocity drift of 30\kmsyr, a mean 
velocity of 8\kms, a velocity gradient of 
$\slantfrac{\partial V}{\partial r}\simeq$4\kms\ mas$^{-1}$ is expected at 220 pc. 
This velocity gradient is larger than that observed in the above maser feature on 
the sky plane ($\sim$1\kms~mas$^{-1}$) and that expected from a steady 
accelerating envelope in the region of the \h2o\ masers estimated by \citet
{bai02}, 

\begin{equation}
\frac{\partial V}{\partial r}\approx \frac{\Delta V}{\Delta r}\simeq
\frac{6\mbox{\kms}}{20 \mbox{AU}}\sim\; 0.1\mbox{\kms\ mas$^{-1}$} . 
\end{equation}

\noindent
If such a large velocity gradient is generated along the line-of-sight, the 
velocity-coherent path is too short to produce strong maser amplification 
occurring in a velocity width of less than 1- 2\kms. Adopting a 
velocity-coherent path 10 times as long as the observed feature size 
($\sim$1 mas), an acceleration motion of less than 1--2\kmsyr\ is 
expected along the line-of-sight. Thus, only large acceleration along 
the sky plane is detectable. 

The present paper is the first approach to directly detect the pulsation-driven 
shock waves, which should be identified in future works by finding 
simultaneously several maser features performing such acceleration motions 
and by elucidating the relation between the occurrence of the acceleration and 
the stellar light curve. 

\section{Summary}

Our monitoring VLBA observations of water masers around RT Vir have revealed 
the 3-D kinematics of water masers in more detail and detected the acceleration 
motions of water maser features. Because of the good angular resolution 
($\simeq$1 mas) and time separations ($\simeq$3 weeks in five epochs), we were 
able to measure 60 proper motions and radial-velocity drifts of maser features. 
Carefully dealing with maser features exhibiting infall towards the star, we 
obtained a spherically-expanding flow model from the 3-D maser kinematics. 
The estimated expansion velocity of $\simeq$8\kms\ is consistent with that 
previously suggested. The estimated outer radius of the maser distribution, 
$\simeq$45 AU, was 2--3 times as large as those previously estimated, except 
for those estimated in the J-Net observations. The estimated velocity field of the 
flow is roughly spherically symmetric, while the maser spatial and velocity 
distribution looks weakly asymmetric. Adopting the above flow model and taking 
into account a statistical parallax, we estimated the distance to RT Vir as to be 
220$\pm$30 pc. We found radial-velocity drifts of maser features of $\leq$2\kms\ 
and speculated that the drifts larger than $>$2\kms\ are due to blending of a few 
maser features within a small region in space and velocity. We also found an 
acceleration motion in the proper motion of a maser feature with a rate of 
33\kmsyr\ and indicating radial acceleration. Such acceleration motions seen both 
in the line-of-sight and on the sky plane are explained by a single common 
phenomenon: passages of pulsation-driven shock waves in the circumstellar 
envelope. The possible variation in the maser expansion size and distribution 
asymmetry with time as well as observed acceleration motions should be 
examined in future observations with time intervals longer than a few cycles of 
the stellar pulsation. 

\acknowledgments

NRAO is a facility of the National Science Foundation, operated under cooperative 
agreement by Associated Universities, Inc. H.~I.\ was financially supported by 
the Research Fellowship of the Japan Society of the Promotion of Science for 
Young Scientist. 


\clearpage
\begin{table*}
{\scriptsize
\caption{Status of the VLBA observations and position-reference feature 
in the proper-motion measurement \label{tab:status}}
\begin{tabular}{llccccrr} \tableline \tableline
& \multicolumn{3}{c}{Observation status} 
& \multicolumn{4}{c}{Position-reference maser feature (RT Vir: I2002 {\it 33})} \\
& \multicolumn{3}{c}{\ \hrulefill \ } & \multicolumn{4}{c}{\ \hrulefill \ } \\
& & & & LSR & Peak & \multicolumn{2}{c}{Position relative to the map origin} \\
& & & Maser & Doppler & intensity & \multicolumn{2}{c}{\ \hrulefill \ } \\
& Date & & features & velocity & at Epoch 1 
& $x=\Delta \alpha\cos\delta$ & $y=\Delta\delta$ \\
Epoch & (1998) & UT span & detected & (\kms) & (Jy beam$^{-1}$) 
& \multicolumn{1}{c}{(mas)} & \multicolumn{1}{c}{(mas)} \\ \tableline
1\ \dotfill & May 11 & 03:00--07:00 & 66 & 17.14 & 29.6 & 83.885 & $-$6.922 \\
2\ \dotfill & May 31 & 01:30--05:30 & 62 & 17.16 & 73.1 & 0.000 & $-$0.069 \\
3\ \dotfill & June 19 & 00:00--04:00 & 53 & 17.10 & 82.3 & $-$0.002 & $-$0.025 \\
4\ \dotfill & July 12 & 22:30--02:30 & 50 & 17.10 & 84.2 & $-$0.015 & 0.059 \\
5\ \dotfill & August 1 & 21:00--01:00 & 53 & 17.10 & 63.5 & $-$0.034 & 0.077 \\ \tableline
\end{tabular}
}
\end{table*}

\clearpage
\begin{table*}
{\tiny
\vspace*{-10mm}
\begin{center}
\caption{Parameters of the water maser features identified by 
proper motion toward RT Vir \label{tab:pmotions}}
\end{center}
\begin{tabular}
{l@{ }r@{ \ }rr@{ \ }r@{ \ }r@{ \ }rr@{ }rr@{ }r r@{ \ }r@{ \ }r@{ \ }r@{ \ }r@{ \ }c} 
\tableline \tableline  
Maser\footnotemark[1] & \multicolumn{2}{c}
{Offset\footnotemark[2]$^{,}$\footnotemark[3]}
 & \multicolumn{4}{c}{Proper motion\footnotemark[3]} 
 & \multicolumn{2}{c}{Rad. motion\footnotemark[4]}
 & \multicolumn{2}{c}{RV drift\footnotemark[5]}
 & \multicolumn{5}{c}{Peak intensity at five epochs} 
 & \\ 
feature & \multicolumn{2}{c}{(mas)} 
 & \multicolumn{4}{c}{(mas yr$^{-1}$)}
 & \multicolumn{2}{c}{(km s$^{-1}$)}
 & \multicolumn{2}{c}{(km s$^{-1}$yr$^{-1}$)}
 & \multicolumn{5}{c}{(Jy beam$^{-1}$)} & Infall\footnotemark[6]  \\ 
 (RT Vir: & \multicolumn{2}{c}{\ \hrulefill \ } 
 & \multicolumn{4}{c}{\hrulefill} 
 & \multicolumn{2}{c}{\hrulefill} 
 & \multicolumn{2}{c}{\hrulefill} 
 & \multicolumn{5}{c}{\hrulefill} & \\ 
 \hspace*{\fill}I2002)  & R.A. & decl. & $\mu_{x}$ & $\sigma \mu_{x}$ 
 & $\mu_{y}$ & $\sigma \mu_{y}$ & $V_{\mbox{z}}$ 
 & $\Delta V_{\mbox{z}}$\footnotemark[7] 
 & $\dot{V_{\mbox{z}}}$ & $\sigma \dot{V_{\mbox{z}}}$\footnotemark[8] 
  & Ep.~1 & Ep.~2 & Ep.~3 & Ep.~4 & Ep.~5 & \\ \tableline 
  1\ \dotfill \  &  $-$89.79 &  7.95 &  1.55 & 0.32 & $-$2.98 & 0.36
 & $-$7.57 & 0.72 & $-$0.02 & 0.35
  & 57.40 &  ...  &  110.33 &  ...  & 12.00 & Infall \\ 
  2\ \dotfill \  &  $-$90.09 &  7.53 &  2.02 & 1.02 & $-$0.26 & 2.18
 & $-$7.53 & 0.71 & $-$0.82 & 1.38
  &  ...  &  ...  &  ...  & 90.08 & 12.41 & Infall \\ 
  3\ \dotfill \  &  $-$63.40 &  8.82 & $-$3.53 & 0.60 &  $-$10.58 & 0.54
 & $-$7.17 & 0.77 &  1.74 & 0.46
  & 24.65 &  ...  & 19.70 & 48.50 &  ... & Infall \\ 
  4\ \dotfill \  &  $-$83.71 &  6.99 &  $-$10.17 & 0.25 &  1.14 & 0.28
 & $-$6.89 & 0.72 & $-$1.64 & 0.31
  &  132.00 & 47.40 &  127.00 & 81.51 & 25.90 & \\ 
  5\ \dotfill \  & $-$112.87 &  $-$14.68 &  $-$10.95 & 0.86 & $-$7.58 & 0.59
 & $-$6.39 & 0.63 &  0.24 & 0.33
  &  4.09 &  6.99 &  4.56 &  ...  &  2.54 & \\ 
  6\ \dotfill \  & $-$110.78 & 26.95 & $-$8.54 & 0.29 & $-$2.08 & 0.56
 & $-$6.26 & 0.39 &  0.41 & 0.45
  & 14.20 & 12.10 &  8.32 &  3.30 &  ... & Infall \\ 
  7\ \dotfill \  & $-$109.81 & 26.29 & $-$7.83 & 0.85 & $-$5.06 & 1.35
 & $-$6.18 & 0.30 & $-$0.91 & 0.74
  &  8.38 &  5.35 &  3.44 &  ...  &  ... & Infall \\ 
  8\ \dotfill \  & $-$110.78 & 27.00 & $-$7.11 &  10.84 & $-$5.04 & 9.27
 & $-$6.17 & 0.44 &  0.73 & 1.45
  & 15.67 & 12.46 &  ...  &  ...  &  ... & Infall \\ 
  9\ \dotfill \  & $-$112.01 & 29.95 & $-$5.62 & 1.24 & $-$0.31 & 0.99
 & $-$6.06 & 0.51 & $-$2.39 & 0.45
  &  7.72 &  5.66 &  4.18 &  2.18 &  ... & Infall \\ 
 10\ \dotfill \  & $-$115.57 &  $-$16.36 &  $-$14.64 & 1.95 & $-$4.85 & 4.77
 & $-$5.99 & 0.50 &  0.87 & 1.38
  &  ...  &  ...  &  ...  &  2.24 &  2.39 & \\ 
 11\ \dotfill \  & $-$110.97 & 27.49 &  $-$10.35 & 0.97 & $-$0.54 & 1.59
 & $-$5.84 & 0.18 & $-$0.84 & 0.74
  &  5.48 &  6.22 &  2.78 &  ...  &  ... & \\ 
 12\ \dotfill \  &  $-$79.21 &  $-$46.69 & $-$8.03 & 0.33 &  $-$10.06 & 0.42
 & $-$5.47 & 0.86 &  0.75 & 0.45
  &  7.29 &  9.04 &  9.50 & 12.11 &  ... & \\ 
 13\ \dotfill \  &  $-$82.47 &  $-$48.46 & $-$1.54 & 3.24 & $-$9.20 & 1.63
 & $-$5.30 & 0.53 &  0.82 & 1.45
  &  2.91 &  1.98 &  ...  &  ...  &  ... & Infall \\ 
 14\ \dotfill \  &  $-$80.38 &  $-$47.83 & $-$0.73 & 2.20 &  $-$11.49 & 1.22
 & $-$5.22 & 0.63 & $-$1.59 & 1.45
  &  4.31 &  4.14 &  ...  &  ...  &  ... & Infall \\ 
 15\ \dotfill \  &  $-$85.89 &  $-$49.26 & $-$5.63 & 0.25 & $-$6.55 & 1.05
 & $-$5.20 & 0.65 &  0.82 & 0.45
  &  ...  &  7.86 &  8.41 &  ...  &  4.98 & \\ 
 16\ \dotfill \  &  $-$86.21 &  $-$48.79 & $-$7.20 & 0.60 & $-$6.70 & 0.24
 & $-$5.14 & 1.01 &  0.53 & 0.31
  & 10.71 & 14.30 & 14.20 &  8.26 & 10.70 & \\ 
 17\ \dotfill \  &  $-$87.32 &  $-$50.13 & $-$6.47 & 0.87 & $-$4.92 & 0.79
 & $-$4.57 & 0.44 & $-$0.02 & 1.38
  &  ...  &  ...  &  ...  &  3.66 &  2.42 & \\ 
 18\ \dotfill \  &  $-$87.59 &  $-$54.85 & $-$6.53 & 0.55 & $-$8.85 & 1.26
 & $-$4.03 & 0.44 & $-$0.39 & 0.46
  &  0.58 &  0.51 &  ...  &  0.38 &  ... & \\ 
 19\ \dotfill \  & $-$108.43 &  $-$52.72 &  $-$12.10 & 1.42 & $-$5.74 & 1.44
 & $-$4.03 & 0.71 & $-$0.86 & 0.66
  &  ...  &  ...  &  0.74 &  ...  &  2.45 & \\ 
 20\ \dotfill \  &  1.35 & $-$9.16 & $-$0.28 & 0.38 & $-$0.13 & 0.62
 & $-$4.03 & 0.62 &  1.78 & 0.31
  &  0.40 &  0.76 &  1.05 &  1.90 &  0.21 & \\ 
 21\ \dotfill \  & $-$105.53 &  $-$49.51 & $-$8.25 & 1.82 & $-$8.48 & 1.83
 & $-$3.97 & 0.53 & $-$0.20 & 0.74
  &  0.59 &  0.73 &  0.68 &  ...  &  ... & \\ 
 22\ \dotfill \  &  $-$87.55 &  $-$53.13 & $-$6.23 & 0.60 & $-$6.42 & 1.45
 & $-$3.97 & 0.66 &  0.22 & 0.45
  &  1.42 &  1.16 &  0.76 &  0.42 &  ... & \\ 
 23\ \dotfill \  &  $-$27.73 &  $-$89.82 &  1.73 & 0.71 & $-$7.42 & 1.22
 & $-$3.68 & 0.31 &  0.93 & 0.45
  &  ...  &  0.38 &  ...  &  0.38 &  0.22 & \\ 
 24\ \dotfill \  &  $-$66.42 &  $-$59.16 & $-$2.38 & 1.52 &  $-$10.97 & 5.06
 & $-$3.66 & 0.74 &  0.33 & 1.45
  &  1.15 &  0.72 &  ...  &  ...  &  ... & \\ 
 25\ \dotfill \  &  $-$27.22 &  $-$88.85 &  1.85 & 2.48 &  $-$16.60 & 3.62
 & $-$3.58 & 0.37 &  1.15 & 1.45
  &  0.42 &  0.48 &  ...  &  ...  &  ... & \\ 
 26\ \dotfill \  &  $-$62.84 &  $-$58.57 & $-$5.38 & 0.76 & $-$6.22 & 0.58
 & $-$3.50 & 0.26 & $-$0.06 & 0.31
  &  1.23 &  1.14 &  0.94 &  0.64 &  0.41 & \\ 
 27\ \dotfill \  &  $-$63.71 &  $-$58.24 & $-$3.48 & 2.71 & $-$7.90 & 5.51
 & $-$3.32 & 0.47 &  3.37 & 1.45
  &  0.51 &  0.51 &  ...  &  ...  &  ... & \\ 
 28\ \dotfill \  &  $-$53.53 &  $-$57.89 & $-$7.33 & 2.21 & $-$2.98 & 2.52
 & $-$3.19 & 0.47 & $-$2.06 & 1.38
  &  ...  &  ...  &  ...  &  0.35 &  0.27 & Infall \\ 
 29\ \dotfill \  &  $-$24.46 &  $-$88.86 & $-$3.01 & 0.69 &  $-$12.58 & 0.70
 & $-$3.14 & 0.45 & $-$0.59 & 0.31
  &  0.42 &  0.50 &  0.52 &  0.54 &  0.35 & \\ 
 30\ \dotfill \  &  $-$48.42 &  $-$58.95 & $-$6.23 & 2.09 & $-$4.43 & 1.56
 & $-$2.66 & 0.23 & $-$1.12 & 1.38
  &  ...  &  ...  &  ...  &  0.27 &  0.15 & \\ 
 31\ \dotfill \  & $-$1.90 & $-$3.14 & $-$0.70 & 0.61 &  1.52 & 1.88
 & $-$2.00 & 0.95 &  0.62 & 0.69
  &  ...  &  8.59 &  6.38 &  3.74 &  ... & \\ 
 32\ \dotfill \  & $-$0.37 & $-$2.81 &  6.03 & 1.07 & $-$2.30 & 1.09
 & $-$1.52 & 0.76 &  4.02 & 0.74
  & 14.30 &  ...  &  6.82 &  ...  &  ... & \\ 
 33\ \dotfill \  &  0.00 &  0.00 &  0.00 & 0.27 &  0.00 & 0.43
 & $-$1.06 & 1.14 & $-$0.23 & 0.31
  & 29.58 & 73.11 & 82.28 & 84.17 & 63.47 & \\ 
 34\ \dotfill \  &  1.26 & 22.67 &  9.60 & 2.81 & $-$1.37 & 4.22
 & $-$0.99 & 0.47 & $-$0.25 & 1.53
  &  ...  &  5.37 &  5.57 &  ...  &  ... & \\ 
 35\ \dotfill \  & $-$0.10 &  $-$23.27 & $-$7.50 & 3.39 & $-$0.15 & 5.93
 & $-$0.99 & 0.50 & $-$1.79 & 1.53
  &  ...  &  4.59 &  5.63 &  ...  &  ... & Infall \\ 
 36\ \dotfill \  &  0.81 &  $-$14.73 &  2.22 & 1.36 & $-$2.18 & 1.56
 & $-$0.96 & 0.42 &  1.73 & 0.74
  &  4.08 &  ...  &  3.11 &  ...  &  ... & \\ 
 37\ \dotfill \  &  1.04 & $-$0.18 & $-$0.30 & 0.55 &  0.65 & 0.70
 & $-$0.76 & 0.79 & $-$0.42 & 0.45
  & 30.95 & 24.80 & 14.66 & 16.32 &  ... & \\ 
 38\ \dotfill \  &  2.91 & $-$3.67 &  2.82 & 1.92 & $-$1.18 & 3.07
 & $-$0.62 & 0.29 & $-$2.03 & 1.53
  &  ...  &  2.19 &  2.86 &  ...  &  ... & \\ 
 39\ \dotfill \  &  0.09 &  0.19 & $-$0.29 & 0.25 & $-$1.33 & 0.25
 & $-$0.62 & 0.29 &  0.00 & 0.69
  &  ...  & 14.80 &  ...  &  9.72 &  ... & \\ 
 40\ \dotfill \  &  1.05 & $-$0.25 & $-$1.22 & 0.21 &  2.07 & 0.23
 & $-$0.62 & 0.67 & $-$1.00 & 0.45
  &  ...  & 20.60 &  ...  & 11.00 &  7.36 & \\ 
 41\ \dotfill \  &  $-$19.18 & 34.53 & $-$4.49 & 0.29 &  4.48 & 0.37
 &  2.68 & 0.45 &  0.15 & 0.31
  &  0.51 &  0.43 &  0.67 &  1.69 &  1.34 & \\ 
 42\ \dotfill \  &  $-$10.56 & 31.90 & $-$2.04 & 0.36 &  4.01 & 1.16
 &  3.07 & 0.50 &  0.09 & 0.74
  & 19.30 &  ...  &  1.70 &  ...  &  ... & \\ 
 43\ \dotfill \  &  $-$11.84 & 34.22 &  0.07 & 0.27 &  2.21 & 0.49
 &  3.26 & 0.34 &  1.58 & 0.45
  &  5.05 &  4.35 &  3.33 &  2.73 &  ... & \\ 
 44\ \dotfill \  &  $-$12.01 & 34.57 & $-$1.59 & 1.03 &  3.40 & 0.91
 &  3.32 & 0.51 &  0.91 & 0.66
  &  ...  &  ...  &  2.22 &  3.47 &  2.28 & \\ 
 45\ \dotfill \  &  $-$10.56 & 31.79 & $-$0.55 & 0.39 & $-$1.17 & 3.31
 &  3.38 & 0.50 & $-$0.49 & 1.45
  & 34.77 & 12.06 &  ...  &  ...  &  ... & Infall \\ 
 46\ \dotfill \  &  $-$10.22 & 30.73 & $-$4.22 & 0.73 &  5.12 & 2.01
 &  3.43 & 0.16 &  2.19 & 0.45
  &  ...  & 12.40 &  2.14 &  3.86  &  4.51 & \\ 
 47\ \dotfill \  &  $-$10.33 & 30.78 & $-$5.49 & 0.53 &  6.37 & 0.76
 &  3.70 & 0.93 &  0.00 & 0.31
  & 50.30 & 36.97 & 22.61 & 11.83 &  7.44 & \\ 
 48\ \dotfill \  & $-$9.97 & 30.55 & $-$2.66 & 0.20 & $-$2.49 & 0.82
 &  3.76 & 0.41 &  1.34 & 0.45
  & 47.30 &  6.58 &  5.26 & ... &  ... & Infall \\ 
 49\ \dotfill \  &  $-$10.56 & 30.91 &  2.51 & 0.96 & $-$7.79 & 2.66
 &  3.77 & 0.60 &  2.17 & 0.74
  & 46.85 & 23.60 &  5.39 &  ...  &  ... & Infall \\ 
 50\ \dotfill \  &  $-$73.88 &  $-$48.96 & $-$7.97 & 0.66 & $-$7.66 & 0.48
 &  3.84 & 0.60 &  1.65 & 0.31
  &  2.04 &  3.74 &  2.73 &  4.38 &  2.16 & \\ 
 51\ \dotfill \  &  $-$77.68 &  $-$44.51 &  $-$10.79 & 0.07 & $-$7.30 & 0.08
 &  4.43 & 0.76 & $-$0.43 & 0.31
  &  5.05 &  6.45 &  8.27 & 20.40 & 27.00 & \\ 
 52\ \dotfill \  &  $-$64.93 &  $-$45.77 & $-$9.97 & 1.46 & $-$6.10 & 1.74
 &  5.33 & 0.44 &  0.97 & 1.26
  &  ...  &  ...  &  2.32 &  4.24 &  ... & Infall \\ 
 53\ \dotfill \  &  $-$63.63 &  $-$42.68 & $-$6.23 & 1.88 & $-$4.24 & 1.86
 &  5.47 & 0.76 &  2.28 & 1.26
  &  ...  &  ...  &  2.79 &  7.21 &  ... & \\ 
 54\ \dotfill \  &  $-$62.61 &  $-$34.55 & $-$4.60 & 0.36 & $-$4.01 & 0.33
 &  5.75 & 1.00 & $-$0.33 & 0.45
  &  131.05 & 40.61 & 11.10 &  6.68 &  ... & Infall \\ 
 55\ \dotfill \  &  $-$62.22 &  $-$49.67 & $-$4.31 & 2.84 & $-$5.94 & 2.01
 &  5.76 & 0.45 & $-$1.17 & 1.45
  &  9.66 &  3.13 &  ...  &  ...  &  ... & \\ 
 56\ \dotfill \  &  $-$32.04 &  $-$30.98 & $-$5.02 & 0.42 & $-$6.84 & 0.44
 &  6.28 & 0.46 &  1.09 & 0.31
  &  0.61 &  0.56 &  ...  &  0.86 &  0.77 & Infall \\ 
 57\ \dotfill \  & 15.22 &  $-$28.06 & $-$2.65 & 1.09 & $-$0.89 & 0.50
 &  7.09 & 0.58 & $-$0.33 & 0.45
  &  1.06 &  0.97 &  0.71 &  0.43 &  ... & Infall \\ 
 58\ \dotfill \  & 15.02 &  $-$18.85 &  3.39 & 0.81 & $-$5.02 & 2.28
 &  7.13 & 0.95 & $-$2.20 & 1.45
  &  2.31 &  0.88 &  ...  &  ...  &  ... & \\ 
 59\ \dotfill \  & 14.20 &  $-$28.45 & $-$0.62 & 1.39 & $-$1.31 & 1.45
 &  7.26 & 0.60 & $-$2.77 & 0.74
  &  0.64 &  0.67 &  0.64 &  ...  &  ... & Infall \\ 
 60\ \dotfill \  & 14.30 &  $-$28.29 & $-$4.81 & 1.18 &  2.13 & 1.53
 &  7.42 & 0.74 & $-$0.72 & 0.66
  &  ...  &  ...  &  0.58 &  0.60 &  0.33 & Infall \\ 
\tableline
\end{tabular}
}

\footnotemark[1] Water maser features detected toward RT Vir. 
The feature is designated as RT Vir:I2002 {\it N}, where {\it N} is 
the ordinal source number given in this column (I2002 
stands for sources found by Imai et al. and listed in 2002). \\
\footnotemark[2] Parameters at the first epoch while the feature was 
being detected. \\
\footnotemark[3] Relative value with respect to the location of the 
position-reference maser feature: RT Vir:I2002 {\it 34}. \\
\footnotemark[4] Relative radial velocity with respect to the assumed 
systemic velocity of \vlsr$=$ 18.2\kms. \\
\footnotemark[5] Secular drift rate of a radial velocity. \\
\footnotemark[6] Infall: exhibiting a negative expansion velocity in the 
model fitting adopting only expanding features. \\
\footnotemark[7] Mean full velocity width of maser feature at half intensity. \\
\footnotemark[8] Uncertainty obtained when assuming a measurement error 
of a radial velocity, which equals to the channel spacing (0.056\kms). 
\end{table*}

\clearpage
\begin{table*}[p]
\caption{Best-fit models for the maser velocity field of 
the RT Vir outflow \label{tab:RTVir-model-fit}}

\begin{center}
\begin{tabular}{lr@{}c@{$\pm$}lr@{}c@{$\pm$}lr@{}c@{$\pm$}lr@{}c@{$\pm$}l}
\hline \hline
& \multicolumn{6}{c}{Including infalling features}
& \multicolumn{6}{c}{Only expanding features} \\
& \multicolumn{6}{c}{\ \hrulefill \ } & \multicolumn{6}{c}{\ \hrulefill \ }\\
\multicolumn{1}{c}{Parameter}& \multicolumn{3}{c}{Step 1\footnotemark[1] } 
& \multicolumn{3}{c}{Step 2\footnotemark[2] } 
& \multicolumn{3}{c}{Step 1\footnotemark[1] } 
& \multicolumn{3}{c}{Step 2\footnotemark[2] } \\
\hline Features finally used & \multicolumn{6}{c}{46} 
& \multicolumn{6}{c}{48} \\ \hline
\multicolumn{13}{c}{Offsets} \\ \hline
\multicolumn{13}{l}{Velocity:} \\
\hspace*{10pt} $V_{\mbox{0x}}$\footnotemark[3] (\kms) \dotfill \ 
&$-$1.9 & & 0.7 & $-$2.6 & & 0.3 & $-$3.2 & & 1.4 & $-$3.5 & & 0.6 \\  
\hspace*{10pt} $V_{\mbox{0y}}$\footnotemark[3] (\kms) \dotfill \ 
& $-$0.7 & & 0.5 & $-$0.4 & & 0.4 & $-$4.3 & & 1.0 & $-$3.8 & & 0.7 \\
\hspace*{10pt} $V_{\mbox{0z}}$\footnotemark[4] (\kms) \dotfill \ 
& \multicolumn{3}{c}{0\footnotemark[5]} & 0.7 & & 0.3 
& \multicolumn{3}{c}{0\footnotemark[5]} & 2.5 & & 0.5 \\
\multicolumn{13}{l}{Position:} \\
\hspace*{10pt} $x_{\mbox{0}}$ (mas) \dotfill \ & $-$39 & & 6 
& $-$57 & & 3 & $-$11 & & 9 & $-$8 & & 5 \\
\hspace*{10pt} $y_{\mbox{0}}$ (mas) \dotfill \  & $-$13 & & 5 
& $-$31 & & 2 & $-$14 & & 6 & $-$15 & & 6 \\
\hline \multicolumn{13}{c}{Velocity field} \\ \hline
\multicolumn{13}{l}{Radial outflow:} \\
\hspace*{10pt} $V_{\mbox{0}}$ (\kms) \dotfill \ 
& \multicolumn{3}{c}{...\footnotemark[6]}& $-$0.1 & & 0.4 
& \multicolumn{3}{c}{...\footnotemark[6]}& 6.6 & & 0.6 \\
\hspace*{10pt} $V_{\mbox{1}}$ (\kms arcsec$^{-\alpha}$) \dotfill \ 
& \multicolumn{3}{c}{...\footnotemark[6]}& 1.5 & & 0.1 
& \multicolumn{3}{c}{...\footnotemark[6]}& 2.0 & & 1.5 \\
\hspace*{10pt} $\alpha$ \dotfill \ 
& \multicolumn{3}{c}{...\footnotemark[6]} & $-$0.40 & & 0.03 
& \multicolumn{3}{c}{...\footnotemark[6]} & 0.51 & & 0.39 \\
Distance $d$ (pc) \dotfill \ & \multicolumn{3}{c}{270\footnotemark[7]} & 87 & & 12 
& \multicolumn{3}{c}{270\footnotemark[7]} & 226 & & 16 \\ \hline
RMS residual $\sqrt{S^{2}}$ \dotfill \ 
& \multicolumn{3}{c}{3.02} & \multicolumn{3}{c}{6.84} 
&\multicolumn{3}{c}{2.33} & \multicolumn{3}{c}{4.33} \\ \hline 
\multicolumn{13}{c}{} \\
\end{tabular}
\end{center}

\footnotemark[1] Assuming independent expansion velocities of maser 
features. \\
\footnotemark[2] Asumming a common expansion velocity field 
expressed by equation (12) of \citet{ima00}. \\
\footnotemark[3] Relative value with respect to the position-reference 
maser feature. \\ 
\footnotemark[4] Relative value with respect to \vlsr$\equiv $ 18.2\kms. \\ 
\footnotemark[5] Step 1 assumes the systemic radial velocity:  
$V_{\mbox{0z}}\equiv $ 0.0\kms. \\ 
\footnotemark[6] The solution determines a radial outflow velocity 
$V_{\mbox{exp}}$(i) independently for each feature with a proper motion. \\
\footnotemark[7] Distance is completely covariant with the $z_{\mbox{i}}$ 
and $V_{\mbox{exp}}$(i) and cannot be determined: $d\equiv$270 pc.
\end{table*}

\clearpage

\clearpage 

\clearpage 

\clearpage

\clearpage

\end{document}